\documentclass[a4paper,11pt]{article}
\usepackage{pos}
\usepackage{euscript,subfigure}

\title{$\varepsilon$-forms for non-planar triangles with elliptic curves at two loops}

\author[a]{Xuhang Jiang}
\author*[b]{Xing Wang}
\author[c]{Li Lin Yang}
\author[a]{Jingbang Zhao}

\affiliation[a]{School of Physics and State Key Laboratory of Nuclear Physics and Technology, Peking University, \\
  Beijing 100871, China}

\affiliation[b]{Technische Universit\"at M\"unchen, TUM School of Natural Sciences, Physik Department,\\
85748 Garching, Germany}

\affiliation[c]{Zhejiang Institute of Modern Physics, School of Physics, Zhejiang University\\
Hangzhou 310027, China}

\emailAdd{xing.wang@tum.de}

\abstract{
In this talk, we discuss how to generalize ideas developed for Banana integrals to two two-loop non-planar triangle Feynman integrals involving elliptic curves, which have non-trivial sub-sectors and whose Picard-Fuchs operators share less symmetry than Banana integrals, to obtain the canonical differential equations and to solve them with suitable boundary conditions.
}

\FullConference{16th International Symposium on Radiative Corrections: Applications of Quantum Field Theory to Phenomenology (
RADCOR2023)\\
28th May - 2nd June, 2023\\
Crieff, Scotland, UK\\}

\begin{document}
\maketitle

\section{Introduction}
\label{sect:intro}
Recently, our ability to calculate Feynman integrals has been boosted dramatically with the help of corresponding geometric information. In particular, perturbative predictions at colliders achieved tremendous success due to the understanding of multiple polylogarithms (MPL) \cite{Goncharov:1998kja, Goncharov:2001iea}, which are closely related to the genus-zero Riemann surface. A lot of Feynman integrals are beyond MPLs, starting from NNLO; see \cite{Bourjaily:2022bwx, Weinzierl:2022eaz} for reviews and references therein. To cut the story short, Feynman integrals can be related to complex manifolds, such as (compact) genus-$n$ Riemann surfaces \cite{Huang:2013kh,Georgoudis:2015hca,Doran:2023yzu} and higher dimensional hypersurfaces like Calabi-Yau manifolds \cite{Bourjaily:2018yfy, Klemm:2019dbm, Bonisch:2020qmm, Pogel:2022ken, Pogel:2022vat, Duhr:2022dxb}. As a powerful tool to calculate dimensional-regularized Feynman integrals, the $\varepsilon$-factorized (canonical) differential equations \cite{Henn:2013pwa} are naturally determined by meromorphic functions on corresponding complex manifolds. The simplest case is the sunrise integral, a two-loop example of so-called Banana integrals, whose geometric object is an elliptic curve, as a genus one Riemann surface \cite{brown2013multiple, Broedel:2017kkb, Bogner:2019lfa}.

Recently, $\varepsilon$-form differential equations to equal-mass Banana integrals have been derived \cite{Pogel:2022vat}. In this talk, we will show how the ideas therein can be generalized to general Feynman integrals, like non-planar triangles shown in Fig~\ref{fig:NPTri}. Unlike Banana integrals, these two cases have non-trivial sub-sector dependence. On top of that, although the geometric objects are elliptic curves, an essential ingredient for the $\varepsilon$-forms in the context of Calabi-Yau manifolds called ``$Y$''-invariant already comes into play for the family (b) in Fig~\ref{fig:NPTri}. The workflow in this talk can be applied to other general Feynman integrals. Limited to the scope of the proceeding, we will not present the $\varepsilon$-form with (uniformly transcendental) boundary conditions or the final results in terms of iterated integrals; we will only show one example of the comparison with numerical package \texttt{AMFlow} \cite{Liu:2017jxz, Liu:2020kpc}. 

\section{Setup}
\label{sect:setup}
\begin{figure}[t!]
  \centering
  \subfigure[]
     {
      \label{subfig:diagrama}
      \includegraphics[width=5cm]{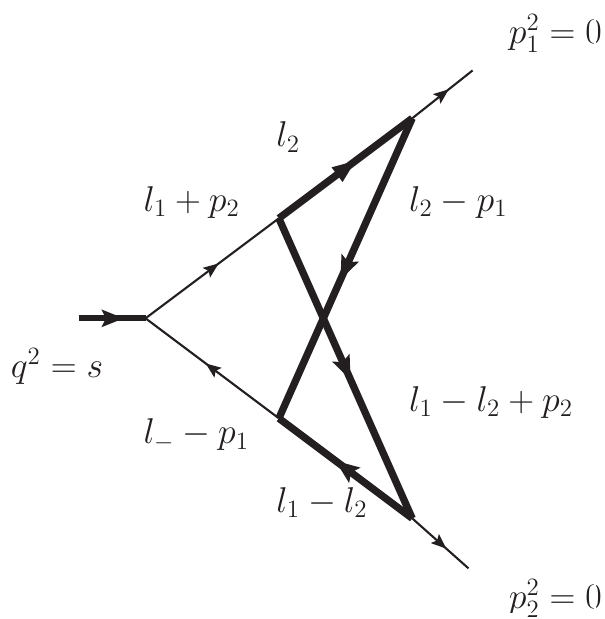}
     }
   \hspace{2em}
   \subfigure[]
     {
      \label{subfig:diagramb}
      \includegraphics[width=5cm]{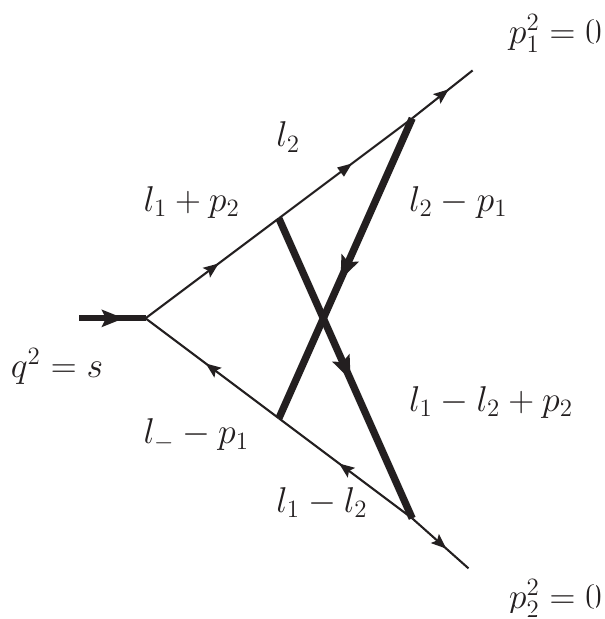}
      }
  \caption{\label{fig:NPTri}Two non-planar triangle integral families at two loops. Family (a) (the left diagram) involves a closed loop with the same mass and was studied in \cite{vonManteuffel:2017hms}. Family (b) (the right diagram) involves two massive propagators.}
\end{figure}

In momentum space, the two Feynman integrals shown in Fig~\ref{fig:NPTri} are represented by
\begin{equation}
	\label{eq:NPTriMomdef}
	\begin{aligned}
		I_{\nu_1\nu_2\cdots \nu_7} = e^{2\varepsilon\gamma_E}(m^2)^{\nu-D}\int\frac{d^D l_1}{i\pi^{D/2}}\int\frac{d^D l_2}{i\pi^{D/2}}\frac{D_7^{-\nu_7}}{D_1^{\nu_1} D_2^{\nu_2} D_3^{\nu_3} D_4^{\nu_4} D_5^{\nu_5} D_6^{\nu_6}} \,,
	\end{aligned}
\end{equation}
with $D=4-2\varepsilon$, $\nu=\nu_1+\cdots+\nu_6+\nu_7$. The propagator denominators are
\begin{equation}\label{eq:nptpd}
	\begin{aligned}
		&D_1 = (l_1-p_1)^2,\,\,D_2 = (l_2-p_1)^2-m^2,\,\,D_3=(l_1+p_2)^2,\,\,\\
		&D_4 = (l_1-l_2+p_2)^2-m^2,\,\,D_5 = (l_1-l_2)^2-\kappa\, m^2,\,\,D_6=l_2^2-\kappa\,m^2,
	\end{aligned}
\end{equation}
where $\kappa=1$ for family (a) and $\kappa=0$ for family (b). We have made the Feynman $i0$-prescription in the propagators implicit. With the pre-factor $(m^2)^{v-D}$, the integrals depend on only one dimensionless variable that we take as
\begin{equation}
	\label{eq:kinevar}
	y = -\frac{m^2}{s} \in \mathbb{R} + i 0 \, ,
\end{equation}
where the infinitesimal imaginary part is determined by the Feynman $i0$ prescription. 

\texttt{Litered}~\cite{Lee:2013mka} and \texttt{Kira}~\cite{Klappert:2020nbg} report 15 master integrals for the family (a), 2 of which are in the top sector and 18 master integrals for the family (b), 3 of which are in the top sector. All master integrals in the sub-sectors can be represented by multiple polylogarithms (MPL)~\cite{Goncharov:1998kja, Goncharov:2001iea}. For the family (a), we choose the same basis in the sub-sector as \cite{vonManteuffel:2017hms, Gorges:2023zgv} to bring the sub-sector into the canonical form. For the family (b), we construct the canonical form in sub-sectors using the method of \cite{Chen:2020uyk, Chen:2022lzr}. Suppose we have the canonical basis for the sub-sectors and let's focus on the top sectors in the following. 

Taking the maximal cut of $I_{1111110}^{(a)}$ and $I_{1111110}^{(b)}$ in the Baikov representation, we obtain the associated elliptic curves:
\begin{equation}
    \label{eq:curves}
	E: \; v^2 = \left(u-u_1\right)\left(u-u_2\right)\left(u-u_3\right)\left(u-u_4\right),
\end{equation}
where the four roots are
\begin{equation}
    \label{eq:roots}
	\begin{aligned}
		{\rm family\,\, (a)}:&\quad u_{1}=0 \,, \quad u_{2}=\frac{1-\sqrt{1-16y}}{2} \,, \quad u_{3}=\frac{1+\sqrt{1-16y}}{2} \,, \quad u_{4}=1, \,\\
		{\rm family\,\, (b)}:&\quad u_1 = -1 \,, \quad u_2 = -\frac{\left(\sqrt{1-8y}+1\right)^2}{4} \,, \quad u_3 = -\frac{\left(\sqrt{1-8y}-1\right)^2}{4} \,, \quad u_4 = 0 \,,
	\end{aligned}
\end{equation}
which contain essential geometric information in the top sectors. The elliptic modulus is then given by
\begin{equation}
    \label{eq:modulusa}
	\begin{aligned}
		{\rm family\,\, (a)}:&\quad k^2=\frac{(u_2-u_1)(u_4-u_3)}{(u_3-u_1)(u_4-u_2)}=\left(\frac{1-\sqrt{1-16y}}{1+\sqrt{1-16y}}\right)^2,\,\\
		{\rm family\,\, (b)}:&\quad k^2 = \frac{(u_2-u_1)(u_4-u_3)}{(u_3-u_1)(u_4-u_2)} = \frac{1-4y-8y^2-\sqrt{1-8y}}{1-4y-8y^2+\sqrt{1-8y}}. 
	\end{aligned}
\end{equation}

Since the family (b) is more complicated than the family (a) and contains all necessary ingredients, we focus on the family (b) from now on.

\section{$\varepsilon$ form }
\label{sect:ansatzb}
This section depicts the workflow to obtain the $\varepsilon$-factorized differential equation for the family (b) in the top sector. We start from the Picard-Fuchs operator in the top sector and work under the maximal cut first and then include sub-sector dependence. The analytic continuation of the modular map follows. 
\subsection{Picard-Fuchs operator}
\label{sect:PFo}
There are three master integrals in the top sector. The first-order coupled differential equation for them is equivalent to the following third-order differential equation for $I^{(b)}_{1111110}$:
\begin{equation}
	\label{eq:PFepsb}
	\begin{aligned}
	L_3^{(\varepsilon)}(y) \, I^{(b)}_{1111110} = \sum_{i=0}^3 r_i(y, \varepsilon)\frac{d^i}{d y^i}   I^{(b)}_{1111110} = \underbrace{\left[\varepsilon^2 \vec{f}_{\rm 2, sub}(y) + \varepsilon^3 \vec{f}_{\rm 3, sub}(y)  \right]\cdot \vec{M}_{\rm sub}}_{R(y,\varepsilon)} \,,
	\end{aligned}
\end{equation}
where the sub-sector integrals $\vec{M}_{\rm sub}$ are the chosen basis mentioned before and $\vec{f}_{\rm 2, sub}(y)$ and $\vec{f}_{\rm 3, sub}(y)$ are rational functions accompanied with several square roots~\cite{Jiang:2023jmk}. The coefficients in the Picard-Fuchs operator are
\begin{equation}
	\label{eq:PFepscoeffsb}
	\begin{aligned}
		r_3(y, \varepsilon) &= 1\,,\quad r_2(y, \varepsilon) = \frac{6 y \big[(1+4y) \varepsilon -1\big]+3}{y (y+1) (8 y-1)}\,,\\
		r_1(y, \varepsilon) &= \frac{\left(16 y^3+12 y^2+4\right) \varepsilon ^2+\left(-24 y^3-18 y^2-12 y\right) \varepsilon +8 y^3+30 y^2+6 y-7}{y^2 (y+1)^2 (8 y-1)}\,,\\
		r_0(y, \varepsilon) &= \frac{1}{y^3 (y+1)^2 (8 y-1)}\left[-(  8 y^2+8 y) \varepsilon^3 -(16 y^3+4 y^2+8 y+8) \varepsilon^2\right.\\
		&\left.\quad+(24 y^3+20 y^2+14 y)\varepsilon -8 y^3-32 y^2-4 y+8\right].
	\end{aligned}
\end{equation}
Although this operator is of order 3, when $\varepsilon=0$, it factorizes into the composition of a second-order operator $L_2^{(0)}(y)$ and a first-order operator $L_1^{(0)}(y)$, found via \texttt{DFactor}~\cite{van2012galois} in \texttt{Maple}:
\begin{equation}
	\label{eq:opfact}
	\begin{aligned}
		L_3^{(0)}(y) &= \underbrace{ \left[\frac{d}{d y}+\frac{8}{8y-1}\right]}_{L_1^{(0)}(y)} \underbrace{\left[\frac{d^2}{d y^2}+\left(\frac{1}{y+1}+\frac{8}{8 y-1}-\frac{3}{y}\right)\frac{d}{d y}+\frac{8 y (y+2)-4}{y^2 (y+1) (8 y-1)}\right]}_{L_2^{(0)}(y)} \, .
	\end{aligned}
\end{equation}
The irreducible second-order operator $L_2^{(0)}(y)$ is associated with the elliptic curve \eqref{eq:curves}. This is the reason that the geometric object in the top sector is an elliptic curve instead of a more complicated object~\footnote{Factorization of the Picard-Fuchs operator is often the case for many Feynman integrals.}. 

Solutions of \eqref{eq:opfact} can be represented by elliptic functions directly from the elliptic curve. However, in general, solutions of irreducible Picard-Fuchs operators with degree greater than 2 are hard to obtain concretely. In this context, one can resort to the Frobenius method \cite{Frobenius+1873+214+235, inceODE:1956, agarwal2008ordinary} to solve for them in a suitable region of $y$, see \cite{Jiang:2023jmk} for details. The upshot is that one can obtain the three solutions of $L_3^{(0)}(y)$ in the neighborhood of $y=0$ as
\begin{equation}
	\label{eq:periodsb}
	\begin{aligned}
		\psi_0(y) &= y^2\left(1 + 2 y + 10 y^2 + 56 y^3 + 346 y^4 + 2252 y^5\right)+\mathcal{O}(y^8),\\
		\psi_1(y) &= \frac{1}{2\pi i} \left[\psi_0\ln y + \frac{y^3}{20}\left(60 + 330 y + 2000 y^2 + 12805 y^3 + 85262 y^4\right) + \mathcal{O}(y^8)\right],\\
		\psi_2(y) &= \frac{1}{(2\pi i)^2}\left[-\psi_0\frac{\ln^2y}{2}+2\pi i\,\psi_1\ln y+\frac{3 y^4}{16}\left(12 + 120 y + 931 y^2 + 6910 y^3\right)+\mathcal{O}(y^8)\right],
	\end{aligned}
\end{equation}
and $\psi_0$ is holomorphic while $\psi_1$ and $\psi_2$ have logarithmic behaviours. One can check that $\psi_0$ and $\psi_1$ are annihilated by $L_2^{(0)}(y)$ and $L_2^{(0)}(y)\psi_2 = \frac{1}{8y-1}$, which is equivalent to \eqref{eq:opfact}. It is natural to use the modular variable and denote the Jacobian from $y$ to $\tau$ as $J$:
\begin{equation}
	\label{eq:tau}
	\tau=\frac{\psi_1}{\psi_0},\quad J\frac{d}{d y}=\frac{1}{2\pi i}\frac{d}{d\tau}
\end{equation}
because $\tau$ characterizes the complex structure on the elliptic curve or, equivalently, the torus. Due to transition invariant $\tau\to\tau +1$ in our context, $q=e^{2\pi i\tau}$ is also useful. By \eqref{eq:periodsb} and \eqref{eq:tau}, the relation between $y$ and $q$ (around $y=0$) is given by
\begin{equation}
	\label{eq:qyb}
	\begin{aligned}
		y(q) &= q - 3 q^2 + 3 q^3 + 5 q^4 - 18 q^5 + 15 q^6 + 24 q^7+\mathcal{O}(q^8) \,,
		\\
		q(y) &= y + 3 y^2 + 15 y^3 + 85 y^4 + 522 y^5 + 3366 y^6 + 22450 y^7 + \mathcal{O}(y^8) \,,
	\end{aligned}
\end{equation}
where the second line can be expressed in terms of Dedekind eta-quotients, and we find that
\begin{equation}
	\label{eq:eta}
	y(\tau) = \frac{\eta(\tau)^3 \, \eta(6\tau)^9}{\eta(2\tau)^3 \, \eta(3\tau)^9} \,.
\end{equation}

Although $L_3^{(0)}(y)$ is not a Calabi-Yau operator, see \cite{Pogel:2022vat, Duhr:2022dxb} in the context of Feynman integrals, we can nevertheless construct special normal forms \cite{Bogner:13} from $\psi_0,\,\psi_1$ and $\psi_2$, from which we define a ``$Y$''-invariant:
\begin{equation}
	\label{eq:Yinv}
	Y= \frac{d^2}{d\tau^2}\frac{\psi_2}{\psi_0}.
\end{equation}
In the three-loop equal-mass Banana case, due to Griffiths transversality or the duality of the Picard-Fuchs operator therein, its ``$Y$''-invariant is trivially constant. In this work, however, the ``$Y$''-invariant is not constant, and interestingly, we find a compact expression for this quantity with the Lambert series:
\begin{equation}
	\label{eq:LambertY}
	Y(\tau) =1+\sum_{n=1}^{\infty} a(n) \frac{q^{n}}{1-q^{n}},\quad \text{with}\quad  a(n)=\left\{ \begin{aligned}
-\left(9 \left\lfloor \frac{n}{6}\right\rfloor+3\right)^{2},& \quad n\equiv 2\,\,({\rm mod}\,6),\\
\left(9 \left\lfloor \frac{n}{6}\right\rfloor+6\right)^{2},&\quad n\equiv 4\,\,({\rm mod}\,6),\\
0,&\quad {\rm otherwise},
\end{aligned} \right.
\end{equation}
where $\lfloor x\rfloor$ stands for the floor function~\footnote{We thank David Broadhurst for encouraging us to pursue this Lambert-series representation and for pointing out that \eqref{eq:LambertY} can be further simplified by factoring out 3 inside the square, such that \eqref{eq:LambertY} has a character $\chi_3(n)=\pm 1, 0$ for $n\equiv\pm 1, 0\,\,({\rm mod} 3)$ respectively.}. 

Equipped with all the above ingredients, we find that the Picard-Fuchs operator can be rewritten as
\begin{equation}
	\label{eq:PFfacb}
	L_3^{(0)}(y) = \frac{\psi_0 Y}{J^3}\Theta_q \frac{1}{Y}\Theta_q^2 \frac{1}{\psi_0}, \quad {\rm with}\quad \Theta_q=q\frac{d}{d q}=\frac{1}{2\pi i}\frac{d}{d\tau}.
\end{equation}
It is easy to check that the right-hand side annihilates $\psi_0$, $\psi_1$, and $\psi_2$ as expected. This can be regarded as a generalization of Banana cases and hints to us about other general Feynman integrals. The form of the Picard-Fuchs operator in $q$ or $\tau$ space illustrates the ansatz for the $\varepsilon$-factorized basis in the top sector:
\begin{equation}
	\label{eq:ansatzb}
	\begin{aligned}
		M_1 &= \varepsilon^4\frac{I^{(b)}_{1111110}}{\psi_0}\,,\\
		M_2 &= \frac{1}{2\pi i\varepsilon}\frac{d}{d \tau} M_1 - F_{11} M_1\,,\\
		M_3 &= \frac{1}{Y}\left[\frac{1}{2\pi i\varepsilon}\frac{d}{d \tau} M_2 - F_{21} M_1 - F_{22}M_2\right] -\vec{g}_{\rm 2, sub}(y)\cdot \vec{M}_{\rm sub} \,,
	\end{aligned}
\end{equation}
where $F_{11},\,F_{21}$ and $F_{22}$ are unknown functions \textit{a priori}. 
We first perform the maximal cut such that we focus on the top sector. The sub-sector dependence will be discussed later on. Under the maximal cut, 
\begin{equation}
	\label{eq:Ansatzde}
	\begin{aligned}
		\frac{1}{2\pi i}\frac{d}{d \tau}
		\begin{pmatrix}
			M_1\\
			M_2\\
			M_3
		\end{pmatrix}^{\text{mc}} 
		=
		\begin{pmatrix}
			\varepsilon\,F_{11} & \varepsilon & 0\\
			\varepsilon\,F_{21} & \varepsilon\,F_{22} & \varepsilon\,Y\\
			A_{31} & A_{32} & A_{33}
		\end{pmatrix}
		\begin{pmatrix}
			M_1\\
			M_2\\
			M_3
		\end{pmatrix}^{\text{mc}} .
	\end{aligned}
\end{equation}
$A_{31},\,A_{32}$ and $A_{33}$ depend on $F_{ij}$'s, $L_3^{(0)}(y)$, and $L_3^{(\varepsilon)}(y)$. Requiring that only terms proportional to $\varepsilon$ survive gives constraints on the rotation coefficients, which can be solved systematically. Readers can refer to \cite{Jiang:2023jmk} for details. The upshot is that $A_{32}=0,\,A_{31}=\varepsilon\frac{64}{27} Y,\, A_{33}=\varepsilon F_{21}$ and that all the letters under the maximal cut are modular forms of $\Gamma_1(6)$, see later discussion.  

\subsection{Sub-sector dependence}
\label{sect:sub}
The sub-sector integrals in Banana cases are tadpoles, and those in the triangles are not tadpoles any longer. If without $\varepsilon^2\vec{f}_{2, \mathrm{sub}}$, then we can set $\vec{g}_{2, \mathrm{sub}}$ in \eqref{eq:ansatzb} to zero. This is the case for the family (a). In the presence of $\varepsilon^2\vec{f}_{2, \mathrm{sub}}$ in the inhomogeneous terms, we can subtract it minimally by requiring
\begin{equation}
	\label{eq:subtractioncondition}
	\begin{aligned}
		\frac{d}{d y} \vec{g}_{\rm 2, sub}(y) = (1-8y) \vec{f}_{\rm 2, sub}(y) \,,
	\end{aligned}
\end{equation}
which is determined by undoing the maximal cut in \eqref{eq:ansatzb} and substituting the inhomogeneous terms $R(\varepsilon, y)$ in \eqref{eq:PFepsb}. Combined with \eqref{eq:Ansatzde} and the canonical basis in sub-sectors, we arrive at the $\varepsilon$-factorized differential equation for all the master integrals in the family (b). 

The presence of several terms in $\varepsilon$ as inhomogeneous contributions for the Picard-Fuchs operator is a generic feature. The subtraction prescription can be generalized straightforwardly. 

\subsection{Analytic continuation}
\label{sect:anac}

The relation between $y$ and $q$ in \eqref{eq:qyb} only holds in a vicinity of $y=0$. In the context of elliptic curves, one can analytically continue the above to a bijection relation valid for the whole kinematic regime, i.e., given a kinematic value $y\in \mathbb{R}+i0$ (with Feynman $i0$ prescription), we can obtain the same value from the following sequence:
\begin{equation}
	\label{eq:bijection}
	y\in \mathbb{R}+i0 \xrightarrow{ \tau(y)}\tau\in\mathbb{H}\cup\{i\infty\}\cup\mathbb{Q} \xrightarrow{y(\tau)}y\in \mathbb{R}+i0,
\end{equation}
where $\mathbb{H}$ is the upper half complex plane with ${\rm Im}\tau>0$ and $\mathbb{Q}$ is the rational field. 

\begin{figure}[hpt!]
  \centering
  \includegraphics[width=0.3\textwidth]{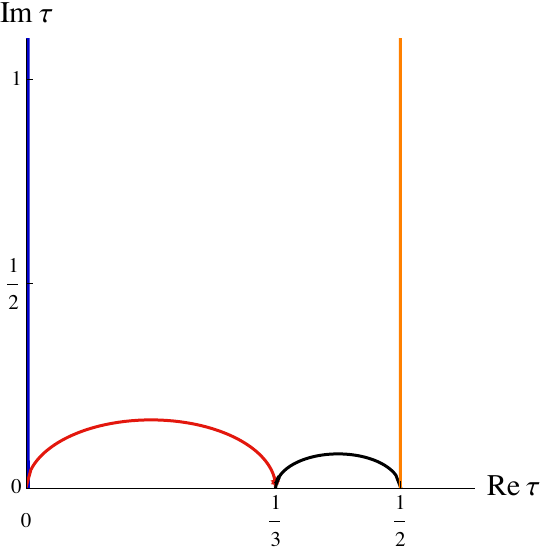}
  \hspace{1cm}
  \includegraphics[width=0.3\textwidth]{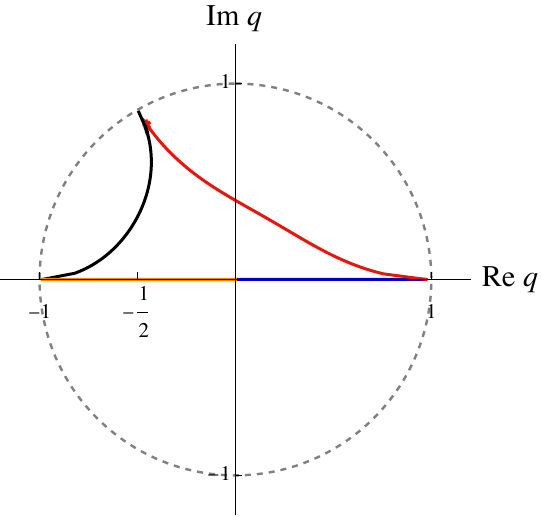}
  \caption{\label{fig:pathb}The behaviors of $\tau$ and $q$ as functions of $y$. The black path corresponds to $y\in (-\infty, -1)$, the orange one corresponds to $y\in [-1,0)$, the blue one corresponds to $y\in [0,\,1/8)$, and the red one corresponds to $y\in [1/8,+\infty) $. }
\end{figure}

For the elliptic curve given by \eqref{eq:curves}, the previous two solutions to construct the modular variable $\psi_0$ and $\psi_1$ are two periods, which are integrals of the only holomorphic differential form on the curve, $du/v$,  on two independent cycles. The integrals turn out to be complete elliptic functions of the first kind. We define two new periods as linear combinations of two complete elliptic integrals:
\begin{equation}
	\label{eq:anacperiodsb}
	\begin{aligned}
		\begin{pmatrix}
			\bar{\psi}_1(y)\\
			\bar{\psi}_0(y)
		\end{pmatrix}=
		\frac{2}{\pi}\frac{y^2}{\sqrt{(u_3-u_1)(u_4-u_2)}} \, \gamma(y) \begin{pmatrix}
			i\,K(1-k^2)\\
			K(k^2)
		\end{pmatrix},
	\end{aligned}
\end{equation}
where the monodromy matrix given by
\begin{equation}
	\label{eq:monodromy}
		\gamma(y) =
		\begin{cases}
		\begin{aligned}
		&\begin{pmatrix}
			1 & 2\\
			0 & 1
		\end{pmatrix}, \quad y < 0 \text{ or } y \geq  \frac{\sqrt{3}-1}{4} \,,
		\\
		&\begin{pmatrix}
			1 & 0\\
			0 & 1
		\end{pmatrix},\quad 0 \leq y < \frac{\sqrt{3}-1}{4} \,,
		\end{aligned}
		\end{cases}
\end{equation}
analytically continue the periods. We refer readers to \cite{Jiang:2023jmk} for details to derive the monodromy matrix and references therein. Performing the series expansion around $y = 0$ and comparing with Eq.~\eqref{eq:periodsb}, we can identify $\psi_0 = \bar{\psi}_0$ and $\psi_1 = \bar{\psi}_1/6$, and
\begin{equation}
	\label{eq:newtaub}
	\tau = \frac{\bar{\psi}_1(y)}{6\bar{\psi}_0(y)} \,,
\end{equation}
plotted in Fig~\ref{fig:pathb}. One can verify that the above function is the inverse map of Eq.~\eqref{eq:eta}, and they satisfy the desired sequence in \eqref{eq:bijection}. 
It is also straightforward to check that the kinematic singular points are mapped to the cusps of the congruence subgroup $\Gamma_1(6)$:
\begin{equation}
   \label{eq:cusps}	
   \tau(y=-1) = \frac{1}{2},\quad \tau(y=0) = i\infty,\quad \tau(y=1/8) = 0,\quad \tau(y=\infty) = \frac{1}{3},
\end{equation}
to which the modular forms in the top sectors belong. It is worth emphasizing that the results obtained in the previous sections don't change under the analytic continuation.

\section{Results}
\begin{figure}[!htp]
  \centering
  \includegraphics[width=0.45\textwidth]{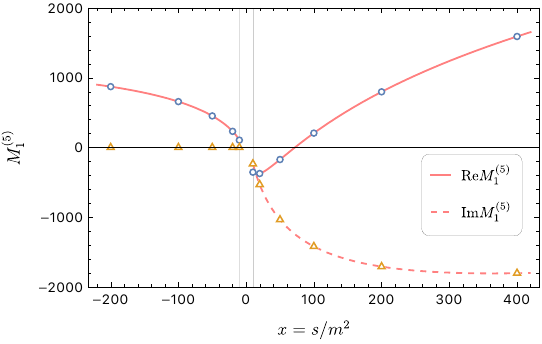}
  \hspace{0.5cm}
  \includegraphics[width=0.45\textwidth]{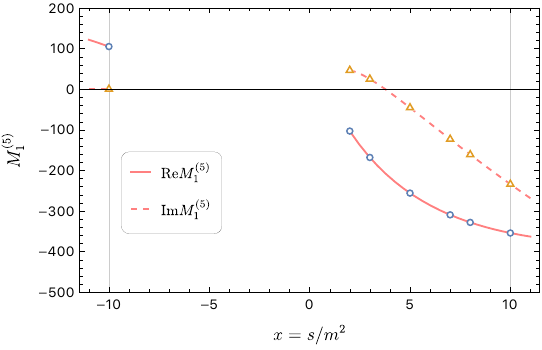}
  \caption{\label{fig:M1w5}Numeric results in family (b) from $q$-expansion (solid and dashed lines) for the weight-5 parts of $M_1=\varepsilon^4 M_1^{(4)}+\varepsilon^5 M_1^{(5)}+\cdots$. Left: a broad range for $|x| > 10$; right: a closer look at $2 < x < 10$. The results are in good agreement with those from \texttt{AMFlow} (circles and triangles).}
\end{figure}

The derived basis in previous sections is not only $\varepsilon$-factorized but also has uniformly transcendental boundary conditions for $y\to 0$, which were calculated by Mellin-Barnes techniques with the help of \texttt{MBTools} \cite{Belitsky:2022gba} and \texttt{XSummer} \cite{Moch:2005uc}. Then, one can solve all the master integrals order by order in $\varepsilon$ in terms of iterated integrals, which can be evaluated numerically very fast by expansion in terms of $q$. Here, we give an example of comparison with \texttt{AMFlow} in Fig \ref{fig:M1w5}. Although all the letters in the top sector are modular forms of $\Gamma_1(6)$, which are holomorphic, the letters in the sub-sector are meromorphic and introduce some poles, such that the $q$-expansion has a finite convergent radius restricted by the nearest pole from the sub-sectors. This is the reason for the blank region in Fig \ref{fig:M1w5}. 

\section{Conclusion}
This talk illustrates how to obtain $\varepsilon$-forms for two non-planar triangle integrals related to elliptic curves. On the one hand, these two integrals have non-trivial sub-sector dependence. We show how to deal with this feature with a subtraction, which can be generalized to other Feynman integrals. On the other hand, the $Y$ invariants developed in the context of Calabi-Yau operators play a role in the family (b), whose Picard-Fuchs operator is not of Calabi-Yau type. We believe this observation can also help obtain $\varepsilon$-forms for other Feynman integrals.

\acknowledgments
This work was partly supported by the National Natural Science Foundation of China under Grant No.~11975030 and 12147103, and the Fundamental Research Funds for the Central Universities. X.W was supported by the Excellence Cluster ORIGINS funded by the Deutsche Forschungsgemeinschaft (DFG, German Research Foundation) under Grant No.~EXC - 2094 - 390783311.

\bibliographystyle{JHEP}
\bibliography{master.bib}
%

\end{document}